# About "SI" Traceability of Micromasses and / or Microforces

Authors: Adriana Vâlcu [1], Dan Mihai Ştefănescu [2]

[1] National Institute of Metrology, [2] Romanian Measurement Society
Bucureşti, Romania

Over the last period, increasing attention has been paid to measurement of small forces which play a more important role in nanotechnology and other significant areas such as MEMS (Micro-Electro-Mechanical Systems) and NEMS (nano-electro-mechanical systems) which can be found into everyday products (mobile phones, MP3 players, PCs, cars). In this respect, the development of mass standards and measurement techniques below the current limit of 1 milligram is vital to provide traceability to the SI for such measurements. In Romania, the Mass laboratory of INM considered it necessary to extend the dissemination of the mass unit below 1 mg, in order to meet current needs. Using the subdivision method and starting from the national prototype kilogram No. 2, all necessary experiments were performed for the first time in Romania to extend mass unit traceability till 100 μg. This extension also supports the provision of mass calibrations for low force measurements. The associated measurement procedure and measurement uncertainty results obtained in the calibration are described. In the article are also presented some of the worldwide methods currently used for measuring small forces.

## 1. Introduction

The lowest traceable force realized by a deadweight standard machine is typically around the level of 1 N, but, in principle, small forces below this limit could be also realized. The National Measurement Institutes (NMIs), motivated by the need for small force standards, have started to explore methods for establishing a hierarchy of SI-traceable force standards at low-force level, consisting of a primary realization, a transfer standard, and methods of dissemination to instruments.

## 2. Methods Used for Force Measurements. A Short Description

The main methods used for the force measurement are [1]:
a) MASS BALANCE, where the unknown force is balanced against a known mass using a digital mass comparator.
The gravimetric calibration by using mass standards is much more accurate (with two orders of magnitude) than by using force measurements, based on the dependence of some electric, magnetic, acoustic or optical parameters variation with the applied load.
b) FORCE BALANCE, i.e. balancing force via a magnet-coil arrangement, called electromagnetic force compensation (EMFC), as applied in PTB and KRISS, or by means of electrostatic force balance (EFB), used in NIST, NPL and CMS;
c) DEFLECTION TYPE TRANSDUCERS measuring the deflection of an elastic element, e.g. piezoresistive cantilever as portable microforce calibration standard.
In the present paper a general scheme, Fig. 1, is proposed for the implementation of micromasses and microforces measurements traceability, starting from various achievements of the NMIs in this area.

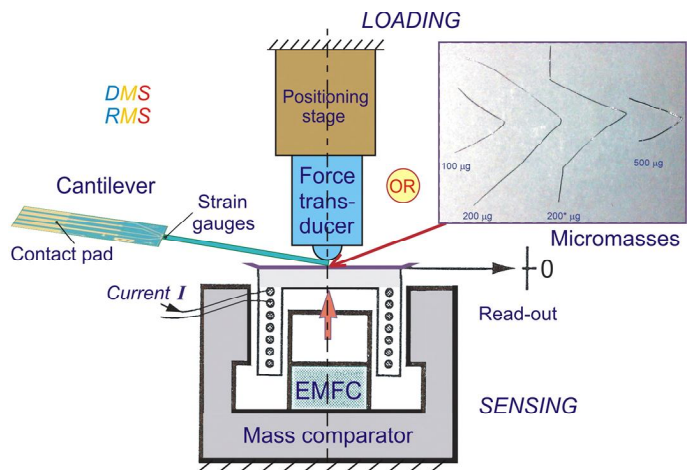

Fig. 1. A general scheme for the implementation of micromasses and microforces measurements traceability

The main idea [2] to develop primary standards based on deadweights is to use a mass balance. Instead of direct application of a mass artifact, force transducers (e.g. interposed piezoresistive cantilevers) are pressed against a mass comparator using a precision positioning stage. Electromagnetic forces generated by a coil and a permanent magnet in the mass comparator counteract the mechanical forces so as to maintain a constant balance position. Then these electromagnetic forces are compared with corresponding traceable deadweights by weighing calibrated mass artifacts on the mass comparator; in this manner, the mechanical forces applied to the transducer can indirectly be compared with the deadweights.

A piezoresistive cantilever could be used as a portable microforce calibration standard [3]. This cantilever-type silicon sensor enables direct calibration immediately before and after a cell probing experiment. Resolution is better than 0.1 μN in a range of (1…50) μN and reproducibility is between 5 and 7 %.

## 3. Weighing Instrument and Micromass Standards Used in Calibration

### 3.1. *Weighing instrument used in calibration*

The weighing instrument used in our research is an UMX 5 balance (Mettler fabrication), Fig. 2, which operates in an EMFC (**E**lectro**M**agnetic **F**orce **C**ompensation) mode, namely the mass of a sample (weighed object) is determined by measuring the force that is exerted by the sample on its support in the gravitational field of the Earth.

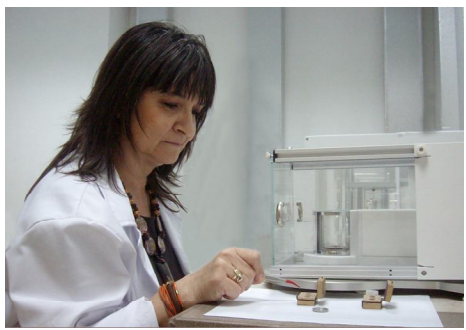

Fig. 2. The weighing instrument, UMX 5 mass comparator, used in calibration

The UMX 5 can be used as mass comparator (for the calibration of the weights), whereas for the next stage, in the low force measurements can be used as a direct weighing instrument. The weighing instrument has the following specifications:

- maximum capacity: 5.1 g;
- readability: 0.000 1 mg;

### 3.2. *Air density measurement equipment*

The mass of an object is obtained by weighing in air. Because the weighing instrument indicates a value that is proportional to the gravitational force on the object reduced by the buoyancy of air, the instrument's indication in general has to be corrected for the buoyancy effect. The value of this correction depends on the density of the object and the density of the air.

The usual method of determining air density is to measure temperature, pressure and humidity and calculate air density using the equation recommended by the Comité International des Poids et Mesures (CIPM) [4].

The mass laboratory is located in a basement and the air conditions are controlled.

### 3.3. *Description of the micromass*

For calibration a set of micromass standards was used, having wire shape (Fig. 3,a) kept in a protection box along with a handling tool (Fig. 3,b). All the weights are made of aluminum alloy. At this moment, this limits the minimum mass of the standard to about 100 µg, any smaller mass being difficult to handle. The set of microstandards consists of the following sequence of nominal values: (5; 2; 2; 1).

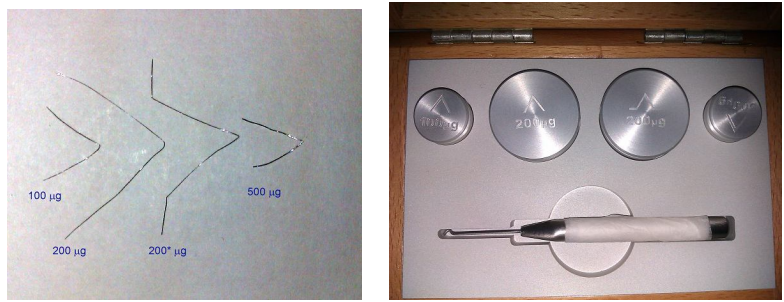

Fig. 3. a) Micromass wire shape; b) Box containing wire micromass

## 4. Measurement Model

### 4.1. *Short description of the mass comparisons*

Generally, for mass comparisons the substitution method is used which allows a simpler balance design and offers more ease of handling and shorter times of measurements.

The substitution method (known also as Borda method) consists in weighing the reference standard A and the test piece B one after another on the same load carrier according to the weighing cycle A-B-B-A, to eliminate possible linear drifts of the balance. The nominal values of the two standards are identical. The indication of the mass comparator is used only for determining the difference between the reference standard A and the test weight B.

In the calibration of mass standards, when the highest accuracy is required, the comparison by subdivision method is mainly used (the calibration of the set weights is performed in itself). When a set of weights is calibrated, one (or more) reference mass standard of a definite nominal value is used for weights with different nominal values. When using only one reference standard, the weighing scheme is an overdetermined system of weighing equations and an appropriate adjustment calculation should be performed in order to avoid propagating errors.

If more than one reference weight is used, the number of equations may be equal to the number of unknown weights. In this case, no adjustment calculation is necessary.

In short, the comparison consists of comparing groups of the same nominal value. These groups may be composed in several variants to have the same nominal value; comparing between them these variants allows a control upon all the performed measurements.

The results of the calibration are obtained with the aid of least squares adjustment which also provides the variance-covariance matrix of the calculated mass values.

### 4.2. *Description of the measurement model*

Using as reference standard a mass of 1 mg, made of aluminum alloy, five micromass standards are calibrated arranging them in twelve possible pairs; from these microweights, as check standard a foil shape micromass standard is used, having nominal mass of 0.1 mg.
The calibration data used are obtained from weighing cycles ABBA for each $y_i$ (which is the weighing comparison according to design matrix "$X$").
The comparison scheme can be represented in matrix form as follow:

$$Y = X\beta + e \tag{1}$$

where
$Y$ (n,1)   is the vector of the n observations (including buoyancy corrections);
$\beta$ (k, 1)   the vector of the k mass values of the standards to be determined;
$X$ (n, k)   design matrix; entries of the design matrix are +1, –1, and 0, according to the role played by each of the parameters (from the vector $\beta$) in each comparison;
$e$ (n,1)   vector of the deviations;
$s_i$ (n,1)   is the vector containing standard deviation of the mean value of each mass difference.
The general mathematical model for "y", corrected for air buoyancy, is:

$$y = \Delta m + (\rho_a - \rho_o)(V_1 - V_2) \tag{2}$$

with:
$\Delta m$   is the difference of balance readings between two weights;
$\rho_o$   1.2 kg m$^{-3}$ the reference air density;
$\rho_a$   air density at the time of the weighing;
$V_1, V_2$   volumes of the weights (or the total volume of each group of weights) involved in a measurement.

$$X = \begin{bmatrix} 1 & 0.5 & 0.2 & 0.2 & 0.1 & 0.1 \\ 1 & -1 & -1 & -1 & -1 & 0 \\ 1 & -1 & -1 & -1 & 0 & -1 \\ 0 & 1 & -1 & -1 & -1 & 0 \\ 0 & 1 & -1 & -1 & 0 & -1 \\ 0 & 0 & 1 & -1 & 1 & -1 \\ 0 & 0 & 1 & -1 & 1 & -1 \\ 0 & 0 & 1 & -1 & -1 & 1 \\ 0 & 0 & 1 & -1 & -1 & 1 \\ 0 & 0 & 1 & 0 & -1 & -1 \\ 0 & 0 & 1 & 0 & -1 & -1 \\ 0 & 0 & 0 & 1 & -1 & -1 \\ 0 & 0 & 0 & 1 & -1 & -1 \end{bmatrix} \quad s_i = \begin{bmatrix} s_1 \\ s_2 \\ s_3 \\ s_4 \\ s_5 \\ s_6 \\ s_7 \\ s_8 \\ s_9 \\ s_{10} \\ s_{11} \\ s_{12} \end{bmatrix} = \begin{bmatrix} 0.00004 \\ 0.00007 \\ 0.00007 \\ 0.00010 \\ 0.00008 \\ 0.00008 \\ 0.00008 \\ 0.00008 \\ 0.00007 \\ 0.00007 \\ 0.00007 \\ 0.00007 \end{bmatrix} mg \quad Y = \begin{bmatrix} y_1 \\ y_2 \\ y_3 \\ y_4 \\ y_5 \\ y_6 \\ y_7 \\ y_8 \\ y_9 \\ y_{10} \\ y_{11} \\ y_{12} \end{bmatrix} = \begin{bmatrix} -0.00062 \\ 0.00579 \\ -0.00076 \\ 0.00556 \\ 0.00674 \\ 0.00674 \\ -0.00556 \\ -0.00556 \\ 0.00781 \\ 0.00781 \\ 0.00723 \\ 0.00723 \end{bmatrix} mg \quad \beta = \begin{bmatrix} \beta_1 \\ \beta_2 \\ \beta_3 \\ \beta_4 \\ \beta_5 \\ \beta_6 \end{bmatrix} = \begin{bmatrix} 1 \wedge \\ 0.5 \wedge \\ 0.2 \wedge \\ 0.2 \Lambda \\ 0.1 \wedge \\ 0.1 \square \end{bmatrix} \tag{3}$$

The first row of the matrix $X$ represents the difference in mass between the +1 and the –1 weight, for example:
1000 - (500 + 200 + 200* + 100) = $y_1$
To estimate the unknown masses of the weights, the least square method was used [5, 6, 7].
The design matrix "$X$" and the vector of observations "$Y$" are transformed in $X'$ and $Y'$ respectively. This transformation is usually performed when the observations are of unequal accuracy (to render them of equal variance). Taking into account the fact that such micromass standards are calibrated and the scale division of the comparator is very small (0.1 µg), any influence which can affect the results should be considered.

$$X' = G \cdot X \quad \text{and} \quad Y' = G \cdot Y \tag{4}$$

$G$ is a diagonal matrix containing the diagonal elements:

$$g_{ii} = (\sigma_0/s_i)^2, \quad i = 1\ldots n \tag{5}$$

and

$\sigma_0$ a normalization factor defined by [5]:

$$\sigma_0^2 = 1/\Sigma(1/s_i^2), \quad i = 1\ldots n. \tag{6}$$

$$X' = \begin{bmatrix} 0.4738 & -0.4738 & -0.4738 & -0.4738 & -0.4738 & 0 \\ 0.3028 & -0.3028 & -0.3028 & -0.3028 & 0 & -0.3028 \\ 0 & 0.2965 & -0.2965 & -0.2965 & -0.2965 & 0 \\ 0 & 0.2055 & -0.2055 & -0.2055 & 0 & -0.2055 \\ 0 & 0 & 0.2495 & -0.2495 & 0.2495 & -0.2495 \\ 0 & 0 & 0.2495 & -0.2495 & 0.2495 & -0.2495 \\ 0 & 0 & 0.2495 & -0.2495 & -0.2495 & 0.2495 \\ 0 & 0 & 0.2495 & -0.2495 & -0.2495 & 0.2495 \\ 0 & 0 & 0.2760 & 0 & -0.2760 & -0.2760 \\ 0 & 0 & 0.2760 & 0 & -0.2760 & -0.2760 \\ 0 & 0 & 0 & 0.2760 & -0.2760 & -0.2760 \\ 0 & 0 & 0 & 0.2760 & -0.2760 & -0.2760 \end{bmatrix} \quad Y' = \begin{bmatrix} -0.00030 \\ 0.00175 \\ -0.00023 \\ 0.00114 \\ 0.00168 \\ 0.00168 \\ -0.00139 \\ -0.00139 \\ 0.00216 \\ 0.00216 \\ 0.00199 \\ 0.00199 \end{bmatrix} \tag{7}$$

The estimates of the unknown masses are calculated, giving the next results:

$$\langle \beta \rangle = (X'^T X')^{-1} X'^T Y' = \begin{bmatrix} 0.0005 \\ 0.0002 \\ 0.0009 \\ 0.0003 \\ -0.0003 \\ -0.0066 \end{bmatrix} mg \tag{8}$$

## 5. Estimating Uncertainty and Consistency of the Results

In evaluating standard uncertainty associated with the results of calibration, the following contributions have to be taken into account:
- type A uncertainty;
- type B uncertainty given by:
- reference standard,
- resolution of the weighing instrument;
- sensitivity of the weighing instrument;
- effect of the air buoyancy.

All these components can be graphically represented in an Ishikawa (Fishbone) diagram, as is shown in the Fig. 4.

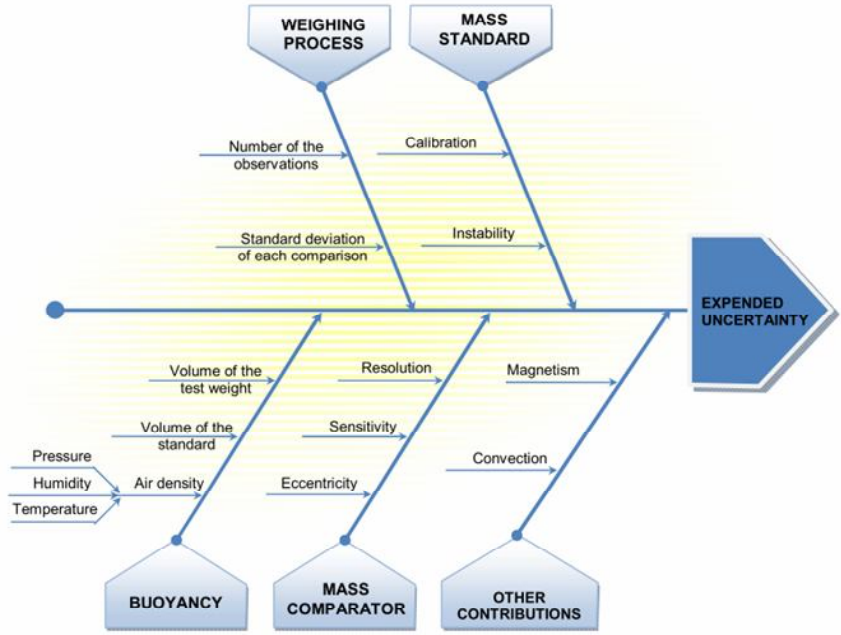

Figure 4. Ishikawa diagram of uncertainty components in micromasses determination

## 5.1. *Type A component*

The standard deviation (uncertainty of type *A*) of a particular unknown weight is given by:

$$u_{A(\beta j)} = s \sqrt{c_{ij}} \qquad (9)$$

where $c_{ij}$ are the diagonal elements of the matrix $(X'^T \cdot X')^{-1}$ and "$s$" is the group standard deviation calculated as follows[5]:

$$s^2 = \{\sum [s_i^2(n_i-1)\cdot g_{ii} + <e_i'>2]\}/f \qquad (10)$$

with
$f$, degrees of freedom", $f = (\sum n_i) - M$;
$n$ = number of weighing equations;
$M$ = the total number of the weights.
If $<y'> = X'\langle\beta\rangle$ are the estimates of the weighted weighing results, the vector of the weighted residuals, $<e'>$ can be obtained from:

$$<e'> = y' - <y'> \qquad (11)$$

## 5.2. *Type B components*

5.2.1. *Uncertainty of the reference standard* comprises: a component from calibration certificate and, another one from its drift $D$ (or stability of standard $u_{stab}$).

$$u_{mcr} = \sqrt{\left(\frac{U}{k}\right)^2 + u_{stab}^2} = \sqrt{\left(\frac{U}{k}\right)^2 + \left(\frac{D}{2\sqrt{3}}\right)^2} \qquad (12)$$

5.2.2. *Uncertainty associated with the display resolution of the balance, $u_{rez}$,* (for electronic balances) is calculated according to the formula [8]:

$$u_{rez} = \sqrt{\frac{d}{2\sqrt{3}} \cdot \sqrt{2}} \qquad (13)$$

### 5.2.3. *Uncertainty due to the sensitivity of the balance*

When the balance is calibrated with a sensitivity weight (or weights) of mass, $m_s$, and standard uncertainty, $u_{(ms)}$, the uncertainty contribution due to sensitivity is [8]:

$$u_s^2 = \Delta m_c^2 \cdot (U_d/2)^2 \tag{14}$$

where $U_d$ is uncertainty of determining the value of the scale interval:

$$U_d = k \cdot u_d = 2d \times \sqrt{\left(\frac{u_{ms}^2}{m_s^2} + \frac{u^2(\Delta I_s)}{\Delta I_s^2}\right)} \tag{15}$$

$\Delta I_s$    the change in the indication of the balance due to the sensitivity weight;
$u(\Delta I_s)$ the uncertainty of $\Delta I_s$;
$\Delta m_c$    the average mass difference between the test weight and the reference weight.

### 5.2.4. *Uncertainty associated with the air buoyancy corrections, $u_b$ is given by [8]:*

$$u_{b\,(\beta j)}^2 = (V_j - V_r h_j)^2 u_{\rho a}^2 + (\rho_a - \rho_o)^2 u_{Vj}^2 + [(\rho_a - \rho_o)^2 - 2(\rho_a - \rho_o)(\rho_{a1} - \rho_o)]u_{Vr}^2 h_j \tag{16}$$

where:
$V_j, V_r$    represents the volume of test weight and reference standard, respectively;
$\rho_a$    air density at the time of the weighing;
$u_{\rho a}$    uncertainty for the air density determined at the time of the weighing, calculated according to CIPM formula;
$\rho o$    1,2 kg·m$^{-3}$ is the reference air density;
$u^2_{Vj}, u^2_{Vr}$ uncertainty of the volume of test weight and one of the reference standards, respectively;
$\rho_{a1}$    air density determined from the previous calibration of the standard;
$h_i$    the ratio between the nominal values of the unknown weights and one of the reference.

### 5.3. *The internal consistency of the results*

From the standard deviation $s_i$, the normalization factor and the group standard deviation are calculated according to equation (6) and (10) giving [5]:
$\sigma_0$ = 0,0000207 mg; $s$ = 0,00002103 mg and $s/\sigma_0$ = 1.02 which shows us that is close to value of 1 and confirms the internal consistency of the observed weighing results.

### 5.4. *Uncertainty budget*

Having all the necessary data, the uncertainty budget (Table I) can be drawn up. This table contains all the components described before.

Table 1. Uncertainty budget

| Uncertainty component | | Standard uncertainty contribution (mg) | | | | | |
|---|---|---|---|---|---|---|---|
| | | 1mg | 500 µg | 200 µg | 200 µg | 100 µg | 100 µg |
| $u_{mcr} \cdot h_j$ | in mg | 0,0006 | 0,00032 | 0,00013 | 0,00013 | 0,00006 | 0,000064 |
| $V_r \cdot h_j$ | in cm³ | 0,00013 | 0,00006 | 0,000025 | 0,000025 | 0,000013 | 0,000013 |
| $u_{Vr} \cdot h_j$ | in cm³ | 2E-06 | 0,000 | 0,000 | 0,000 | 0,000 | 0,000 |
| $V_j$ | in cm³ | | 0,00019 | 0,00007 | 0,00007 | 0,00004 | 0,000037 |
| $u_{Vj}$ | in cm³ | | 1,37E-06 | 5,49E-07 | 5,49E-07 | 2,74E-07 | 0,000000 |
| $\rho_a$ | mg/cm³ | | 1,1716 | 1,1716 | 1,1716 | 1,1716 | 1,1716 |
| $u_{\rho a}$ | mg/cm³ | | 0,0006 | 0,0006 | 0,0006 | 0,0006 | 0,000606 |
| $(V_j - V_r \cdot h_j)^2 u_\rho$ | in mg | | 5,49E-15 | 8,79E-16 | 8,79E-16 | 2,20E-16 | 2,20E-16 |
| $(\rho_a - \rho_o)^2 u^2_{Vj}$ | in mg | | 1,52E-15 | 2,44E-16 | 2,44E-16 | 6,09E-17 | 6,09E-17 |
| $[(\rho_a - \rho_o)^2 - 2(\rho_a - \rho_o)(\rho_{a1} - \rho_o)] u^2_{Vr} \cdot h_j$ | | | -1,62E-15 | -2,60E-16 | -2,60E-16 | -6,50E-17 | -6,50E-17 |
| $u_b^2$ | in mg | | 5,39E-15 | 8,63E-16 | 8,63E-16 | 2,16E-16 | 2,16E-16 |
| $u_{res}$ | in mg | | 0,000041 | 0,000041 | 0,000041 | 0,000041 | 0,000041 |
| $u_s$ | in mg | | 0 | 0 | 0 | 0 | 0 |
| $u_{conv}$ | in mg | | 0 | 0 | 0 | 0 | 0 |
| $u_{ecc}$ | in mg | | 0,0001 | 0,0001 | 0,0001 | 0,0001 | 0,0001 |
| $u_{ma}$ | in mg | | 0 | 0 | 0 | 0 | 0 |
| $u_{ba}$ | in mg | | 0,000108 | 0,000108 | 0,000108 | 0,000108 | 0,000108 |
| $u_A$ | in mg | | 0,000035 | 0,000024 | 0,000024 | 0,000024 | 0,000026 |
| $u_c =$ | | | 0,00034 | 0,00017 | 0,00017 | 0,00013 | 0,00013 |
| $U =$ | | | 0,00068 | 0,00034 | 0,00034 | 0,00026 | 0,00026 |

## 6. Conclusion

A wide range of derived SI units (Flow, Volume, Force, Pressure, Humidity and Density) is based on mass determination for traceability. An improvement in the realization and dissemination of the mass scale leads to an improvement of measurement uncertainty in these areas.

This way, the aim of the work described here is to extend the mass unit traceability till 100 µg which can lead to an extending of force unit traceability till 1µN.